# Optical and Terahertz Response of Carbon Nanostructures

Arvind Singh and Sunil Kumar[*]

*Femtosecond Spectroscopy and Nonlinear Photonics Laboratory,
Department of Physics, Indian Institute of Technology Delhi, New Delhi 110016, India*
*\*Corresponding Author: kumarsunil@physics.iitd.ac.in*

## Abstract

In the last three decades or so, we have witnessed an extraordinary progress in the research and technology of carbon-based nanomaterials. Among the peculiar highlights are the discoveries of fullerene, the carbon nanotubes and the magnificent simple scotch tape exfoliated graphene. The unique photophysical properties of these different allotropic forms of the nanocarbon have opened up vast application possibilities in many fields of science and technology, with particular emphasis on optoelectronics and photonics. A prerequisite for many of these applications is a thorough understanding of the nature of the elementary and coupled excitations and also various dynamical processes involving them. Here, we present an overview of the recent excitement with the carbon nanostructures, in particular, the quantum dots, nanotubes and graphene. We discuss some of their very interesting properties investigated through optical and THz spectroscopic tools. At optical frequencies, the light emitting properties, the nonlinearities and ultrafast response have been presented, while, the low-energy response has been considered in terms of studies obtained by using THz time-domain spectroscopy. Finally, we conclude with some of the future prospects on the photophysics of carbon nanosystems in realistic applications.

## 1. Introduction

Carbon, as the main building block of organic compounds, is one of the most important chemical elements essential for all living organisms. Much of the discussion on nanotechnology perspectives has been centered around carbon-based nanostructures in the last three decades due to the popularity of fullerenes, nanotubes, and unprecedented success of graphene. The interest in low-dimensional carbon materials has been exponentially growing since then, particularly, because of their exceptional electrical, chemical, thermal, and optical properties[1, 2]. Carbon-based nanomaterials, in particular, carbon nanotubes (CNTs) and graphene, provide a variety of new opportunities for fundamental and applied research across many disciplines of science. A huge amount of data, understanding and new discoveries can be found in the literature on these materials, including those due to the optical techniques in a large spectrum spanning all the way from ultraviolet (UV) to terahertz (THz) frequencies. In this dimension of research on the carbon nanostructure, here, we have reviewed a few interesting topics that indicate the current excitement amongst researchers.

Consisting of carbon chains arranged in unique geometric patterns, nanocarbons, i.e., the carbon nanostructures, can conduct electricity, absorb, and emit light, exhibit interesting photonic and optoelectronic properties[3-5]. Some of these are made of $sp^2$-hybridized carbon atoms and can be classified according to their dimensionality. As shown in Fig. 1, fullerenes are zero-dimensional (0D), CNTs are one-dimensional (1D), and graphene is two-dimensional (2D) form of the nanocarbon. The bulk graphite is three-dimensional (3D) carbon material made of graphene layers stacked together via weak van der Waal forces. Since their discovery in 1985, the bulky molecules of carbon, i.e., fullerenes[6], have enriched the palette of carbon chemistry and have become one of the most important building blocks in the emerging field of nanoscience and nanotechnology. The abundant form of fullerene, $C_{60}$, displays an exceptional redox chemistry[7]. Also, the ability to transport charges in three dimensions, these properties together, render $C_{60}$ and its derivatives, suitable for areas such as photosynthetic reaction mimics, photovoltaics, and catalysis[8]. Within the last three decades, notable breakthroughs in terms of basic research on electron transfer, have paved the way for applications, including organic solar cells and novel electron donor-acceptor conjugates as well as hybrids[9]. Other than fullerenes, CNTs are at a crossing point between molecules and solid-state materials. CNTs can be found in single-walled, double-walled, and multi-walled nanotube forms. The properties of a CNT can be determined by its chiral indices (m,n). Both, the multi-walled nanotubes (MWNTs) and single-walled tubes (SWNTs) were discovered by Iijima[10, 11] in 1991 and 1993, respectively. CNTs are of interest for electronic applications because of their extraordinary structural properties (extremely high aspect ratios), mechanical properties (elastic modulus of ~1TPa and tensile strength of ~100GPa), and electronic properties (small band gaps in semiconducting SWNTs, high charge mobilities, current capacities, and heat capacities in metallic SWNTs). A major drawback is, however, the intrinsic polydispersity of each carbon nanotube type in any sample that is synthesized. Important breakthroughs in terms of producing monodisperse carbon nanotube samples have lately been reported, but only on small scales[12]. Continued efforts are going on in the direction of achieving controlled growth of samples, which contain only one type of nanotubes having the predetermined chiral index (m,n), diameter and so on.

Graphene as a monolayer of carbon atoms in the form of a strict 2D hexagonal lattice structure, provides a lot of fascinating physics. It has a unique band structure, which is fundamentally different from anything that was known before its discovery. The linear dispersion in the low-energy regime, where the electrons behave like relativistic massless particles, give rise to many extraordinary properties. In the vicinity of the Dirac points in the energy space, the electrons in graphene are called Dirac fermions and can be described by a Dirac-like equation of relativistic quantum mechanics. For a long time, it was presumed that freestanding graphene can never be found in nature, since strictly two-dimensional crystals could not exist due to the thermodynamically instability[13]. However, in 2004, Novoselov and Geim demonstrated for the first time that graphene flakes can be isolated from graphite by a mechanical exfoliation technique[4]. They used a simple scotch-tape method for this purpose. Soon after, a number of studies started reporting some unusual and unique physical properties of graphene[14]. Novosolov and Geim were awarded the most prestigious Nobel Prize in Physics in year 2010, only 6 years after the reporting of their discovery. Since then, graphene has been a subject of intense research for its remarkable electronic, mechanical, optical, and chemical properties and potential for future applications[14]. The existence of an unusual quantum Hall effect was also discovered in graphene[15, 16].

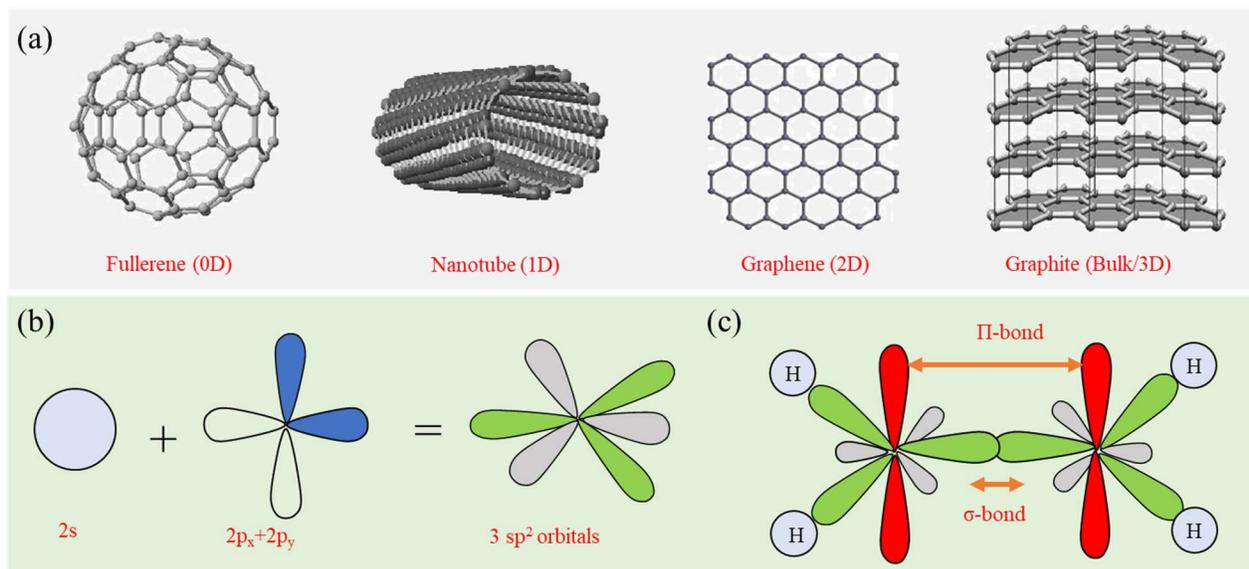

**Figure 1:** (a) Structures of representative nanocarbons classified according to their dimensionality. Fullerenes ($C_{60}$), carbon nanotubes, graphene and bulk graphite represent 0D, 1D, 2D and 3D nanostructures, respectively. (b) The intermixing of atomic orbitals (2s, $2p_x$, $2p_y$) of the carbon atoms gives rise to three new $sp^2$ hybrid molecular orbitals. (c) σ-bonds are in the xy-plane and π-bonds formed by fourth valence electron in the $2p_z$ orbital, are perpendicular to the xy-plane.

A few applications have been described in Fig. 2 arising from the cutting edge research efforts by researchers on nanocarbons. Quantum confinement effect in nanosized carbon dots/particles provides tunable light emitting properties. In addition to having large surface to volume ratio, some other unique virtues of carbon nanodots, such as low toxicity, chemical stability, and fluorescence properties, make them an exceptional vehicle for fluorescence labelling and imaging in cost-effective medical diagnosis techniques. Carbon dots can aid diagnosis by delivering information regarding the type, size, and position of tumours in humans. Counting on their therapeutic functions, fairly good biocompatibility[17], weak interactions with proteins, resistance to swelling and photobleaching, easy clearance from the body, and immune system evasion[18-20], carbon dots can help effectively deliver and controllably release drugs to the specific parts of the body. The nonlinear optical properties of quantum-confined nanocarbon structures are interesting both as spectroscopic tools from the basic research point of view and as passive mode-locking devices for ultrashort lasers to generate picosecond (ps) and femtosecond (fs) laser pulses[21]. Proper ultrafast saturation behavior and carrier energy relaxation properties are essential for the later[22]. Mou *et al.*,[23] used CNTs to mode-lock an erbium-doped soliton fiber laser, while Xie *et al.*,[24] employed graphene to mode-lock a solid-state laser. Kono *et al.*,[25] demonstrated a THz polarizer that was built with stacks of aligned SWNTs. The device exhibits ideal broadband THz properties, 99.9% degree of polarization and extinction ratios of ∼$10^{-3}$ (or 30 dB) from ∼0.4 THz to 2.2 THz.

There have been a large number of optical and THz spectroscopic studies on these materials, performed in the weak-excitation and quasi-equilibrium regime with ultrashort time resolution as well, to investigate and assess their performance characteristics for fast optoelectronic applications. In the subsequent sections below, we provide a review from the literature on light emitting properties of carbon dots and nanotubes, and take it further to discuss a few studies on the ultrafast optical processes and THz properties of CNTs and graphene performed using ultrafast time-resolved optical pump-probe and THz time-domain spectroscopic techniques. In Section 2, starting from the typical light emitting properties of carbon dots of varying size, we have also included fluorescence behavior of SWNTs and temperature-dependent photoluminescence from carbon dots in a film. In Section 3, we discuss nonlinear optical and ultrafast processes in graphene and carbon nanotubes. Schematics of the typical experimental arrangements for such studies are also provided. Static THz conductivity investigated by time-domain spectroscopy and optically induced transient THz conductivity in graphene and CNTs, investigated by time-resolved optical pump-THz probe spectroscopy have been discussed in Section 4.

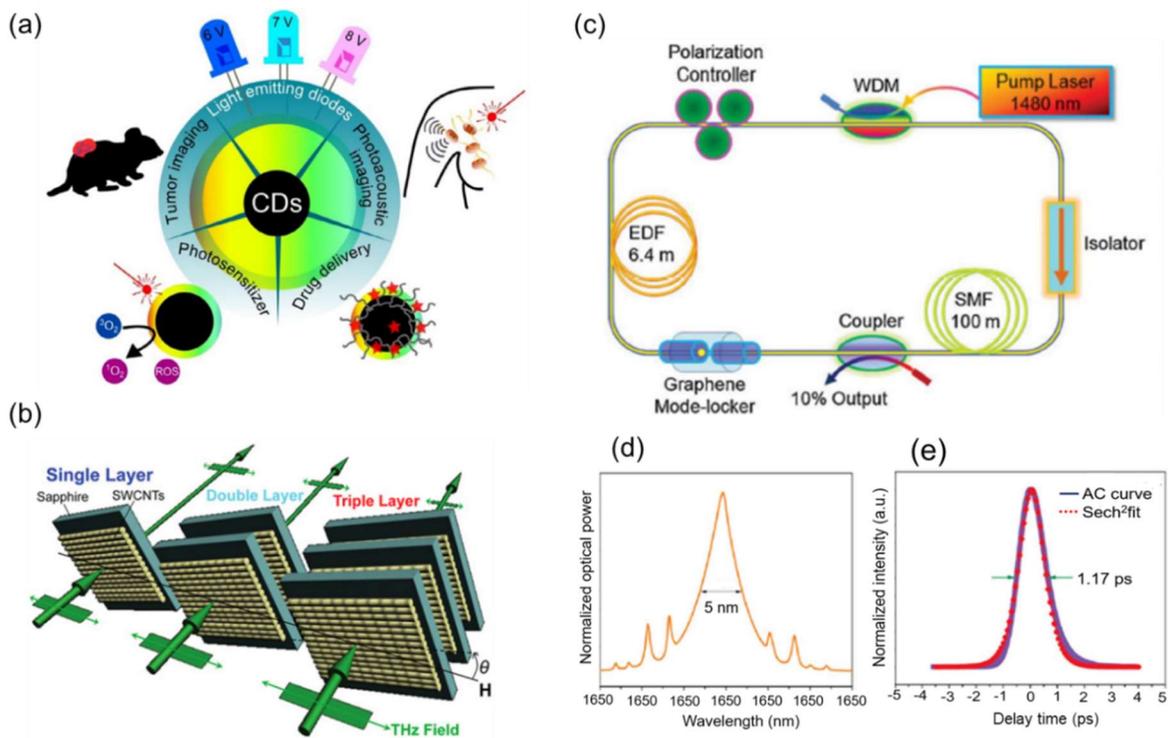

**Figure 2:** (a) Applications of carbon-based nanosized dots in the light emitting devices, biological and medical fields. Adapted from ref. [20]. (b) Scheme showing the use of multiple SWNT films to produce high-performance THz polarizers [25]. (c) An ultrafast laser cavity mode-locked by use of a graphene based saturable absorber, (d) spectrum of the mode-locked laser output, and (e) intensity autocorrelation trace of the laser pulse fitted with $Sech^2$ function to determine its full width at half maximum duration. Adapted from ref. [21].

## 2. Light emitting properties of carbon nanostructures

Reducing the dimensions and/or symmetry of a bulk material will typically change its optical properties[26]. As mentioned earlier, all nanosized carbon materials except nanodiamonds consist primarily of $sp^2$-hybridized carbon atoms and thus possess π-electrons that can undergo optical transitions in a broad range. Reduced dimensionality of the bulk carbon material intrinsically introduces structural disorders, for instance, by functionalization, to create regions of $sp^3$-hybridized carbon regions surrounding embedded aromatic domains. This produces additional states, making the material photo luminescent, most of the times[27-29]. Nanosized allotropes of carbon have always attracted considerable interest because of their diverse light emitting properties depending on their crystal structure, size, morphology, and chemical functionalization. Carbon based dots (carbon nanodots and graphene nanodots) were first isolated by Xu *et al.*, in 2004 as a byproduct of crude arc discharge soot purification[30]. Since then, a huge library of synthetic strategies for preparing carbon-based dots have been developed to obtain different surface functionalities, tunability in the colloidal stability and solubility of the product material[31, 32]. Below, we present the photoluminescence (PL) properties, their control, and origin of PL in different forms of carbon nanostructures. The key parameters addressed include, shape, size, structural type (single wall versus bundles and multi-chiral carbon nanotubes), and the roles of doping, sample temperature, and surface functional groups in the PL behavior of nanocarbons.

The broad emission of carbon based nanodots (CDs) is frequently considered to be a integrated effect of $sp^3$ hybridized surface-located emitters and core-embedded $sp^2$ aromatic chains[33]. For example, the PL spectra of CDs prepared by dehydration of acetic acid with $P_2O_5$ were deconvoluted using two bands that were assigned to emissions from $sp^2$ clusters (core) and COOH-induced surface states, respectively (Fig. 3a)[33]. Similarly, Wei and Qiu [34] observed convoluted blue and green emission peaks and assigned the green one to the presence of surface defects caused by the introduction of oxygen functional groups. Hola *et al.*,[35] observed that the PL of gallate derived CDs functionalized with ester groups on their surfaces was redshifted when the esters were converted into carboxylic acids by hydrolysis. Graphene nanodots (GNDs) were also reported to exhibit red-shifting of their PL upon increasing surface functionalization because of charge transfer between the GNDs and their functional groups[36]. In fact, different emission colors can be obtained interestingly from single solution of CDs by using appropriate excitation wavelengths or bandpass filters[37]. Varying nanosized graphene oxide fractions isolated by density gradient separation show similar optical properties[38], which is later supported by Vinci and Ferrer[39], where, they also found no clear relationship between particle size and emission color. In contrast, Yang *et al.*,[40] and Li *et al.*,[41] demonstrated size dependent emission colors in different fractions of crystalline CDs obtained hydrothermally from glucose and electrochemically from a graphite rod, respectively (Fig. 3b). In these cases, the emissions became increasingly red-shifted as the size of the CDs increased, in a similar manner to that observed for semiconductor quantum dots.

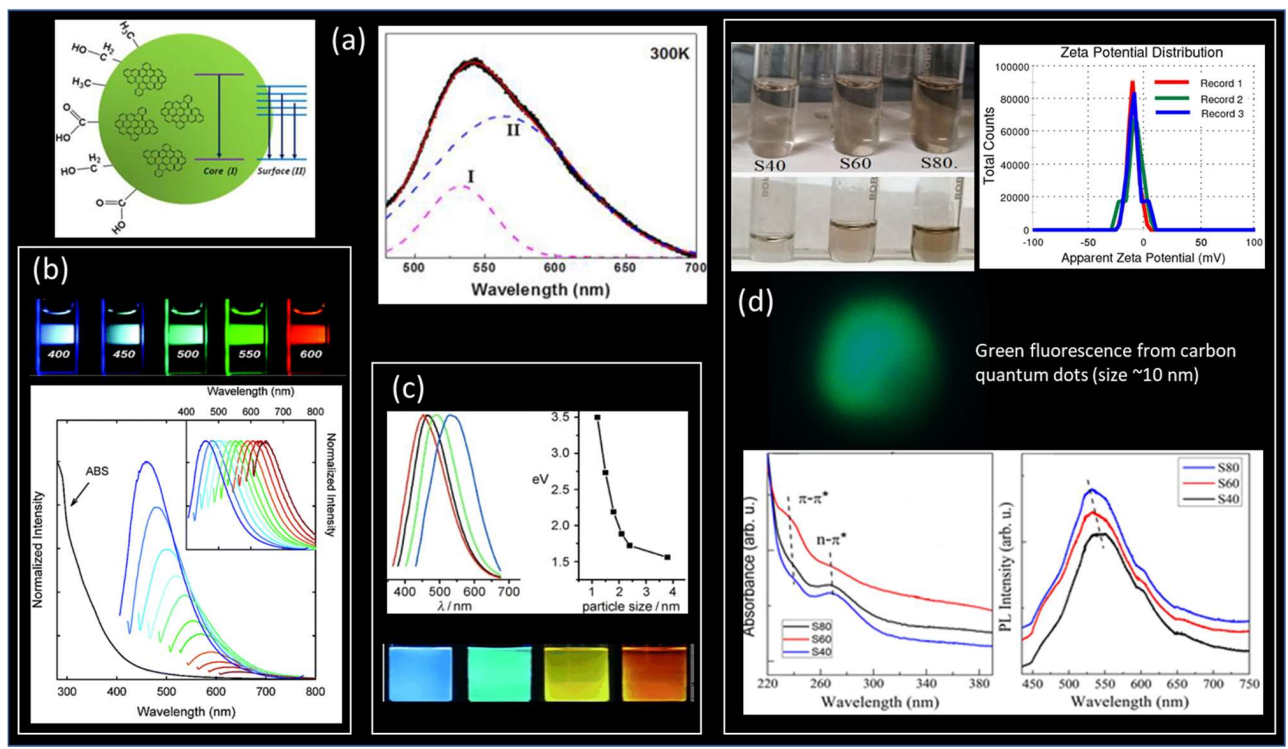

**Figure 3:** Light emitting properties of CDs. (a) Schematic of core- and surface-related energy states (left panel) and the assigned spectral bands in the PL (right panel) for a typical few nanometric size carbon nanoparticle. Adapted from ref. [33]. (b) Top: Fluorescent PEG1500N-attached CDs in water prepared by laser ablation when excited at the indicated wavelengths. Bottom: Absorption and emission spectra of similar CDs passivated with PPEI-EI, in water. The inset shows the dots' normalized PL spectra. Adapted from ref. [37]. (c) PL spectra and images of the samples consisting of CQDs, when illuminated by 365 nm laser light. The PL peak photon energy (eV) vs CQDs size (nm) [41]. (d) Green emission from stable water suspension of carbon quantum dots (CQDs). Top left: Optical images of the CQDs water suspensions as synthesized by nanosecond laser ablation (top row) and after four months (bottom row) for S40, S60 and S80 samples. Please see text for the sample details. Top right: Zeta potential results from three consecutive measurements on S40 sample. Bottom left: UV–Visible absorption spectra, and Bottom right: Photoluminescence spectra at 325 nm laser excitation of CQDs in the S40, S60 and S80 samples. The dashed line is drawn to mark the spectral shift from one sample to the other. Adapted from ref. [42].

Solution stability is one of the central problems in the carbon-based dots synthesis. Recently, we have shown that by nanosecond (ns) laser irradiation of graphitic microparticles in water, a highly stable water suspension of luminescent carbon quantum dots (CQDs) of high purity can be generated[42]. Figure 3(d) shows the optical images of CQDs water suspensions at the time of preparation (top row) and after four months (bottom row). The sizes of the quantum dots are tuned in the range from 5 nm to 15 nm by varying the laser irradiation time from 40 min (S40) to 60min (S60). A highly dispersed and stable suspension of CQDs with no agglomeration is evidently visible. Quantum confinement is strongly evident from the optical absorption and photoluminescence spectra (see bottom part of Fig. 3d). Interestingly, it was observed that the quantum dots are self-functionalized during the fragmentation process. Fourier transform infrared (FTIR) spectrum shows the presence of unsaturated -CO bond at the surface of the quantum dots resulting in the development of similar charges. The unsaturated -CO groups provide a net negative charge on the CQDs surface which helps in stability of the suspensions. Results from the Zeta potential measurements on S40 sample are presented in the top right part of Fig. 3(d). It is found to be −13 mV. Thus, the development of similar charges on CQDs surface leads to the repulsive columbic force between them that eventually prevents agglomeration of the quantum dots leading to a stable water suspension. The formation of the functional groups on the surface of the CQDs is crucial for achieving stable suspension either in water or in any other chemical solvent[43].

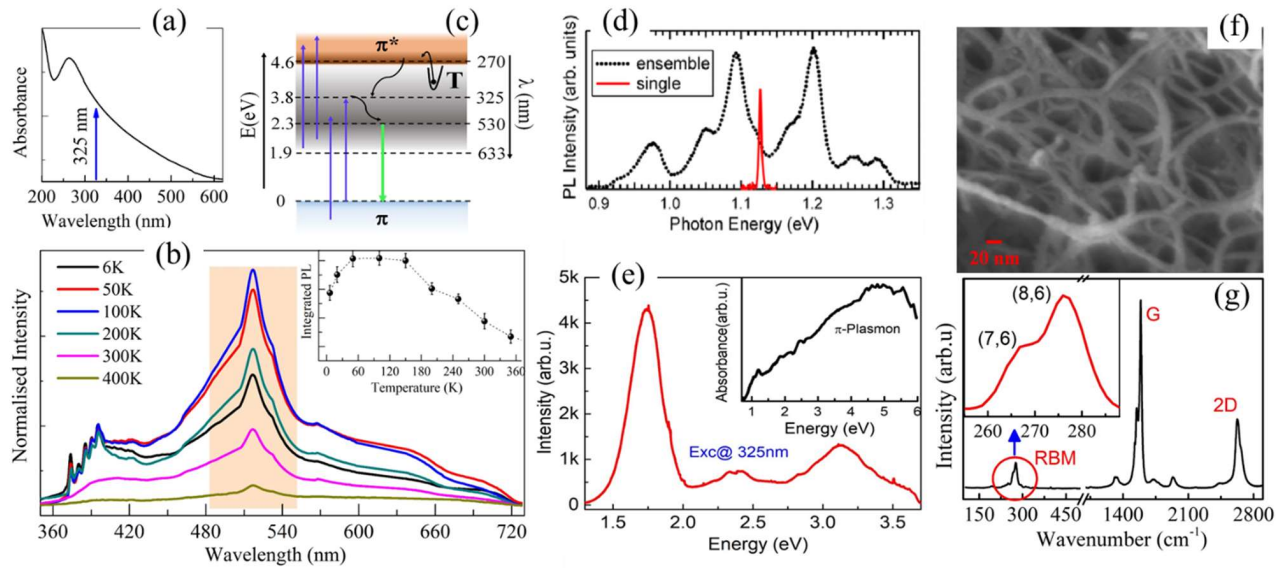

**Figure 4:** (a-c) Optical properties of carbon nanoparticles, CNPs film. (a) UV-Visible absorption spectrum showing a maximum at ∼270 nm due to the π–π* transition and a broad tail extending into the visible range due to the n–π* transition. (b) PL spectra taken at temperatures varying from 6 K to 400 K following laser excitation at 325 nm. Inset: The unusual temperature dependence of the PL intensity maximizing around 100 K seen from integrated PL intensity at 530 nm. (c) Schematic of the energy structure and possible electronic transitions to describe the nature of the optical absorption and PL spectra. Trap states represented by T in the joint configuration space are incorporated within the forbidden gap, which are thermally activated to enhance the PL at an intermediate temperature below the room temperature[44]. (d-f) Optical and morphological properties of carbon nanotubes. (d) Comparison between the PL spectrum of an individual suspended (7, 6) SWNT and ensemble micelle-wrapped SWNTs dispersed in gelatin[45]. (e) PL spectrum taken at 325 nm (3.8 eV) excitation wavelength. Inset: Optical absorbance spectrum. (f) FESEM image of SWNTs (scale bar = 20 nm), showing randomly entangled bundles of SWNTs having an average bundle diameter of ∼10 nm. (g) Raman spectrum showing characteristic RBM and G-bands of SWNTs. Inset: Zoomed-in view for the RBM features[46].

For investigating further, the type and nature of the defects present in the carbon dots, temperature-dependent photoluminescence studies should be carried out. For this, we used a carbon nanoparticles (CNPs) film that was developed directly on a quartz substrate by nanosecond laser ablation of microcrystals of graphitic particles (see Figs. 4(a-c)). Other than oxygenated impurities that are responsible for the luminescent properties of the amorphous carbon nanoparticle films, we find clear signatures of shallow and deep level localized or trap states in the temperature-dependent PL behavior of the film[44]. During the PL measurements, following the laser excitation at 325 nm, immediately the weakly bounded carriers at the impurity levels jump to the conduction band leaving behind vacancies for new photocarriers to be excited from lowest bonding $sp^2$ band (valence band) in a continuous manner. Both of these have been depicted by upward vertical arrows in the schematic of Fig. 4(c). The carriers excited in such a way can recombine through two pathways either (i) directly from mid-gap defect states or (ii) via first relaxing toward lowest energy antibonding $sp^2$ states (in the conduction band) and subsequently to the vacancies created in the impurity states. Also, the electrons from the high lying impurity states can decay to lower impurity states, provided vacancies are found, and then recombine with holes in the valence band through PL emission. The process indicated by the downward green arrow for the PL in Fig. 4(c) is related to the recombination of carriers from $sp^3$ impurity states with the holes in the bonding $sp^2$ states. With the increasing sample temperature, the PL intensity in Fig. 4(b) first increases and then decreases in an unusual manner. This is highlighted by the pink shaded region in the spectra shown in Fig. 4(b). The actual behavior of the temperature-dependent PL is shown in the inset of Fig. 4(b), where the integrated PL intensity has been plotted as a function of sample temperature. Clearly, two temperature regimes occur where the PL kinetics is fundamentally different. The anomalous temperature dependence of the PL cannot be well explained by invoking just the $sp^3$ impurity states only, rather another process marked by T in Fig. 4(b) is also necessary to interpret the unusual temperature effect. In the low temperature regime, i.e., below 100 K, the PL intensity continues growing up with the increasing temperature; however, above 100 K, strong reduction in PL intensity takes over. Upon photoexcitation with 325 nm laser, the significant density of photogenerated carriers get trapped within the shallow and deep level trap states, which were formed in abundance due to the structural defect's formation during the laser ablation process. With the increasing sample temperature, the trapped carriers are thermally activated into the conduction band and migrate toward the radiative $sp^3$ impurity states leading to an enhancement in the PL intensity. This is consistent with the observations from photoconductivity measurements discussed in the same paper, where, the photocurrent decay time-constant continuously decreases with temperature for both laser excitations because the carriers overcome the barriers (traps) with the increasing temperatures. The reduction in the PL intensity above 100 K in Fig. 4(b) signifies thermal quenching of PL due to the dominating role of a large number of phonons mediated non-radiative decay channels over the counter effect due to the activation of trapped carriers. The PL intensity almost vanishes at ~400 K. At a given sample temperature, enhancement in the PL intensity from two same type of material samples having different amounts of defect densities is termed as nonthermal quenching, whereas a pure thermal effect leading to quenching of the PL intensity in the same type of material is due to thermal activation of nonradiative channels in the material. The latter is the prime reason for the observed temperature-dependent PL behavior.

The light emission mechanism in carbon nanotubes is mainly governed by strong excitonic effects. This is due to the 1D quantum confinement effect that makes the binding energy of excitons in CNTs sufficiently high. Consequently, CNTs possess stable excitons even at temperatures beyond than room temperature. The inhomogeneous effects in ensemble SWNTs broadened the PL peaks. The PL spectrum of a pristine, single suspended SWNT directly grown on a patterned groove and ensemble micelle-wrapped SWNTs dispersed in gelatin is shown in Figure 4(d) for the comparison[45]. The PL peak from the single suspended nanotube is quite sharp. In contrast, the PL peaks from the ensemble SWNTs embedded in gelatin have considerably broad widths, suggesting inhomogeneous broadening effects in the gelatin matrix. It has been well discussed in literature that the PL emission energy considerably varies depending on the environment that surrounds the SWNTs[47]. Very recently, we have also seen a very strong PL from SWNTs network of nanotube bundles (see Figs. 4(e-g)). We observed a strong emission feature at ~1.7 eV is followed by a broader emission band at ~2.4 eV in the bi-chiral SWNTs bundles, which are related to the $E_{22}$ transition energies of the two contributing SWNTs species[46]. The bi-chirality of the SWNTs in the thin film was confirmed through radial breathing mode (RBM) features present in Raman spectra taken at 785 nm excitation. The much broader spectral feature centered at ~3.1 eV in the PL spectrum is due to the contributions from $E_{ii}$ manifolds in the tubes. Notably, the strongest PL peak at ~1.7 eV due to the $E_{22}$ transitions is a clear indicator that (8,6) nanotubes are majorly present in bi-chiral SWNTs thin film sample.

## 3. Nonlinear optical processes and ultrafast non-equilibrium carrier dynamics in graphene and carbon nanotubes

There is considerable interest in finding novel materials having large yet fast optical nonlinearities. For applications in all-optical ultrafast switching and sensor protection, one needs to investigate the concerned materials for both the nonlinear optical absorption (NLA) and refraction (NLR) properties. NLA and NLR, although, having different spectral properties always coexist as they both result from the same physical mechanism. Due to the simplicity of the popular z-scan technique as well as the interpretation of the related data, it has gained rapid acceptance by the nonlinear optics community as a standard technique for determining the NLA and NLR properties of homogeneous material structures through open and closed aperture configurations, respectively. The third-order nonlinearities are responsible for the variation of refractive and absorptive properties of isotropic media and the propagation of intense light through the concerned structures. The physical processes that give rise to NLA and NLR can include various possibilities. Some of these are the thermal refraction, the ultrafast bound electronic processes, optically induced excited state processes, Kerr effect, etc. Ultrafast processes include multi-photon absorption and stimulated Raman scattering and AC-Stark effects[48]. Excited-state nonlinearities can be caused by a variety of physical processes including absorption saturation [49], excited-state absorption and free-carrier absorption as well as defect and color center formation[50]. The above processes can lead to increased transmittance with increasing irradiance (e.g., due to saturation absorption) or decreased transmittance (e.g., due to multi-photon absorption, excited-state absorption). A typical schematic of the z-scan experimental setup is shown in Fig. 5(a). For closed and open aperture configuration, one needs to simply insert or remove an aperture in front of the detector. For open aperture configuration, it is necessary that all the light transmitted through the sample is collected on the photodetector, hence use of a lens between the collection lens and the photodetector is recommended. The transmittance of the sample through the aperture is monitored in the far field as a function of the position, z, of the nonlinear sample from the focal point of the focusing lens. The required sample scan range in the z-direction depends on the beam parameters and the sample thickness. A critical parameter is the diffraction length, $z_0$, of the focused beam defined as $\pi w_0^2/\lambda$ for a Gaussian beam, where, $w_0$ is the focal spot size (half-width at the $1/e^2$ of the maximum).

In many practical situations where considerable laser power fluctuations may occur during the scan, a reference beam onto another photodetector can also be used to monitor and normalize the transmittance. The reference arm can be further modified and made identical to the other nonlinear arm for eliminating the possible spatial beam fluctuations. In the linear regime, i.e., when the sample has been placed far away from the focus in closed aperture z-scan configuration, the size of the aperture is meant by its transmittance, S, Commonly, 0.1<S<0.5 has been used for studying nonlinear refraction. Obviously, the S = 1 case corresponds to collecting all the transmitted light and provides the measurement of nonlinear absorption.

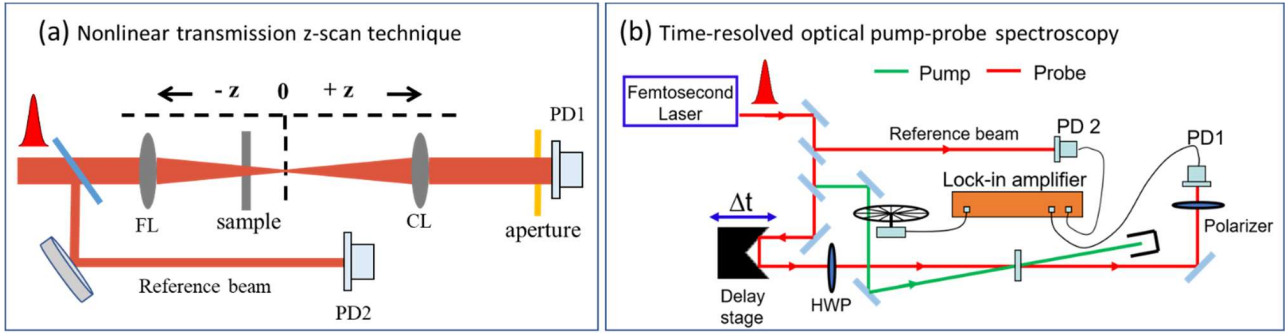

**Figure 5:** Typical experimental arrangements for carrying out studies for nonlinear optical properties and non-equilibrium dynamics. (a) Single beam z-scan technique and (b) time-resolved pump-probe spectroscopy. FL: focusing lens; CL: collimating lens, PD1, PD2: photodiodes, HWP: half wave plate.

Probing ultrafast non-equilibrium dynamical processes in materials taking place at ps and fs time scales, is an arduous task as even the fastest photodetectors available today are quite slow for that matter. In fact, the fastest speed of the electronic photodetectors is fundamentally limited and with them response times in sub-ps cannot be achieved. Thus, for access to the relevant timescales (sub-ns) reliably, alternate nonelectrical methods are needed. One such popular method is called ultrafast time-resolved pump-probe spectroscopy. Of course, ultrafast lasers providing pulses of sub-ps duration are required at the desired wavelengths. Currently, ultrafast lasers providing sub-10 fs laser pulses anywhere from near UV to much into infrared are available or such pulses can be generated from a driver laser by nonlinear mixing processes in suitable optical crystals. Ultimately, the resolution of the ultrafast processes in the material under study are limited to the pulse duration. In a typical pump-probe spectroscopy setting as shown in Fig. 5(b), a high-intensity optical pump pulse perturbs the material system from equilibrium and a time-delayed weak probe pulse measures the photoinduced change in either the transmission or reflection of the sample at different delays between the strong pump pulse and the weak probe pulse. The pump and probe pulses can be taken either from the same laser or two different lasers depending on the requirement, however, a zero-delay point must be ensured before trying out the experiments with the later one. The setup comprises of two important components, i.e., a computer controlled linear translation stage for introducing precise time-delays between the pump and probe pulses and another one, a mechanical chopper for modulating the pump beam so that the pump-induced changes in the probe pulse can be measured in a lock-in amplifier at the modulation frequency. Pump-probe spectroscopy can be performed in two modalities, on in which both are at the same wavelengths and the other when pump and probe wavelengths are different. To improve the sensitivity of the detection, one needs to use appropriate filters and polarizers so that direct pump scattering into the probe detector is avoided. For an absorbing medium, the pump induced modification in the carrier density is responsible for the observed transmission or reflection changes and hence, carrier dynamics can be captured from the temporal evolution of the corresponding signal. Pump probe spectroscopy has an advantage over the time-resolved fluorescence spectroscopy in that the changes of the absorption are sensitive to both the changes in the ground state and the excited state as compared to the fluorescence spectroscopy which measures changes only from the radiative transitions in the sample under investigation.

### 3.1 Ultrafast nonlinear processes in graphene: Saturable absorption

The optical nonlinearity is typically only observed at very high light intensities, such as that provided by pulsed lasers. Graphene shows remarkable optoelectronic properties and optical nonlinearities in a broad spectral range with ultrafast response times. Various nonlinear effects such as saturable absorption (SA), two photon absorption (TPA), four wave mixing (FWM) and optical limiting or reverse saturation absorption (RSA) can be observed in graphene. Fine-tuning of the tailorable nonlinear optical properties of graphene-based material systems can be achieved by rational modification of the chemical structure via oxidation and functionalization and altering the electronic structure of the system. In saturable absorbers, the absorption of light decreases with increasing incident light intensity and hence ultrafast saturation absorption property can be useful for mode-locking in ultrafast lasers. The ultrafast saturation absorption and electronic process in graphene is discussed through Fig. 6(a-f). Reduced graphene oxide (GO) flakes behave in these properties as much as the pristine graphene and hence the time-resolved pump-probe data and the nonlinear transmission data from z-scan are presented in Figs. 6(c-f) from a chemically synthesized graphene oxide (GO) sample in the suspension form so that most of the sample interacting with the excitation light is the individual flakes. As indicated in Fig. 6(a), at a sufficiently high incident intensity, electrons get accumulated in the higher energy states if the rate at which they decay back to the ground state is insufficient. This causes saturation absorption before the ground state can be completely depleted. At high excitation densities, the photo-generated carriers increase in concentration (much larger than the intrinsic electron and hole carrier densities in graphene at room temperature) and cause the states near the edges of the conduction and valence bands to fill [51], blocking further absorption and imparting transparency to light at photon energies just above the band-edge[52]. Due to the Pauli blocking process, saturation absorption or the absorption bleaching is achieved in graphene[53].

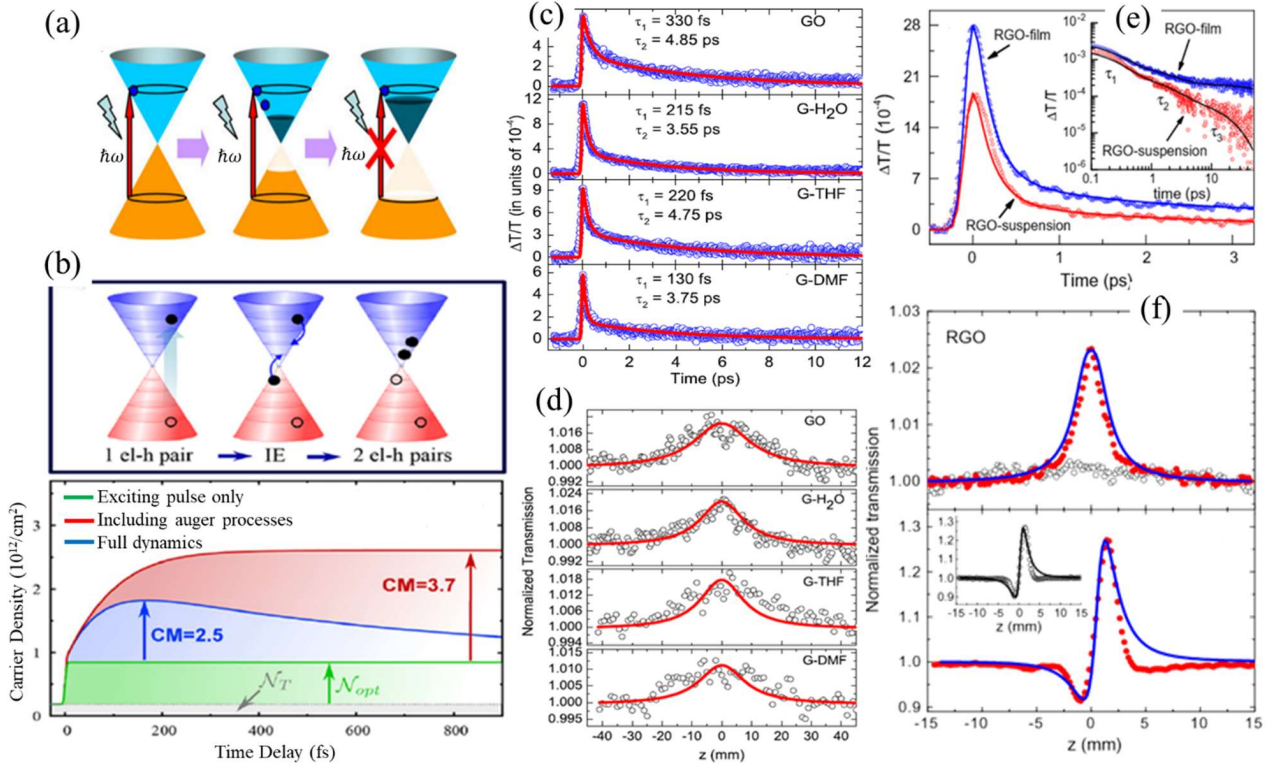

**Figure 6:** (a) Schematic of excitation process responsible for SA behavior of graphene[52]. (b) Carrier multiplication process in graphene. Top: Schematic illustration of the impact excitation multiplying the optically excited electron-hole pairs through internal scattering. Bottom: Temporal evolution of the carrier density, N(t) considering only carrier-light interaction (green line), only Coulomb scattering including Auger processes (red), and full dynamics including carrier-phonon channels (blue line). Here, N(t), $N_{opt}$ and $N_T$ are the total, the purely optically excited, and the thermal carrier density, respectively. Figure adapted from ref. [54]. (c) Transient differential transmission spectra as a function of probe delay in the degenerate pump-probe experiment at 790 nm (1.57 eV), and (d) open aperture z-scan data of the graphene suspensions in different liquids as mentioned[49]. (e) Femtosecond differential probe transmission data for RGO (film and water suspension) with 3.2 eV pump and 1.57 eV nm probe pulses. The inset shows the signals in log–log scale clearly showing three distinct regions characterized by three-time constants. (f) Open and close aperture z-scan results for the RGO-suspension. For reference, same measurements performed on water as reference are shown by open circles. Theoretical fits shown by continuous thick curves determine the nonlinear coefficients. GO: graphene oxide, RGO: reduced graphene oxide. Adapted from ref. [55].

Figures 6(b), 6(c) and 6(e) represent the ultrafast carrier excitation and relaxation dynamics in graphene. Study of the carrier relaxation pathways is important for determining the feasibility of materials for fast electronic and photonic devices. For example, in a mode-locked laser, the shortest pulse duration is approximately limited to the carrier recovery time in the saturable absorber used inside for passive mode-locking[56]. For understanding the carrier dynamics in graphene, one follows from the thermal distribution, optical excitation, carrier thermalization, and carrier cooling processes. Briefly, (i) the starting point is initial carrier temperature mediated thermal fermi distribution. (ii) Next, a non-equilibrium carrier distribution is generated following excitation by an excitation ultrashort light pulse. This generates a highly anisotropic non-equilibrium distribution corresponding to the excitation energy in the density of states belong to conduction band for electrons and valence band for holes. (iii) The non-equilibrium electrons are redistributed to energetically lower states via combined carrier-carrier and carrier phonon scattering resulting into a thermalized carrier distribution within first tens of fs. Same holds true for the hole dynamics in the valence band. (iv) Finally, phonon-induced scattering redistributes energy from the excited carrier system to the lattice on a picosecond timescale resulting in a narrowing of the carrier distribution. Eventually, the initial thermal Fermi distribution is reached after complete energy relaxation. Carrier-phonon scattering processes are involved in both the thermalization and the carrier cooling. High-resolution pump-probe experiments have found two distinct decay times with $\tau_1$ typically in the range of 100 fs and $\tau_2$ typically around 1 ps representing thermalization and cooling times[49, 57]. Similar numbers can be seen in the pump-induced probe transmission changes from the graphene suspension samples presented in Fig. 6(c). In case of films casted on a substrate, the time constants vary slightly (Fig. 6(e)) and also there is a new and much longer third relaxation component which is observed due to different nature of the sample in the film.

Carrier thermalization and carrier cooling discussed above are typical relaxation steps during the carrier dynamics in any material. There are differences in the time scales, but the qualitative behaviour is similar. However, graphene exhibits specific ultrafast phenomena of significant carrier multiplication (CM) in its relaxation dynamics[54]. Carrier multiplication is the generation of multiple electron-hole pairs via internal scattering which is a many-particle phenomenon[58]. Auger scattering, which provides the collinear Coulomb-induced interband relaxation channel is the underlying physical mechanism in CM (top part in Fig. 6b). In contrast to all other Coulomb processes, Auger scattering changes the charge carrier density consisting of electrons in the conduction band and holes in the valence band[59]. To investigate CM phenomenon in graphene, Malic et. al., monitored the temporal evolution of the carrier density by sequentially switching on different scattering channels in a controlled manner. As represented in Fig. 6(b) (bottom part), initially an optical ultrashort pulse (FWHM~10 fs) characterized by a pump fluence of 2.3 $\mu$Jcm$^{-2}$, with excitation energy of 1.5 eV, is applied. Accounting only for carrier-light interaction, the applied ultrashort pulse excites electrons from the valence into the conduction band resulting in an optically induced carrier density, which remains constant even after the excitation pulse is switched off (green line). Taking into account Coulomb-induced scattering channels including Auger processes, a strong increase in the carrier density can be even after the optical pulse has been switched off (red line), hence, a CM factor of 3.7 is noted in the case of a purely Coulomb-driven dynamics[60]. However, phonon-induced recombination counteracts the effect and reduces the CM factor to 2.5. This carrier multiplication has the potential to increase the responsivity in photodetectors as well as power conversion efficiency in photovoltaics[61]. Carrier multiplication is a general phenomenon observed recently also in perovskite materials, however linear band structure of graphene is favourable for process[62]. Auger-induced carrier multiplication in graphene was theoretically predicted initially [59, 60] and confirmed later experimentally[63, 64].

Saturation absorption arising from band filling effects in graphene occurs at a wide range of energies from visible to infrared. Following the pump excitation, spreading in the carrier distribution[65] takes place over a wide range, both above and below the states optically coupled by the pump. The so generated hot carrier population causes absorption saturation not only at the pump photon energy but also at probe photon energies different from the pump[66]. The magnitude of saturation absorption and hence the change in probe transmission is usually smaller for the nondegenerate case as seen for RGO: $(\Delta T/T)_{max}$ ~ $6.2 \times 10^{-3}$ in degenerate pump–probe and $1.9 \times 10^{-3}$ in non-degenerate pump–probe experiments[67]. The nonlinear optical coefficient determined from open and close aperture z-scan experiments as shown in Figs. 6(d) and 6(f) are found to be similar for graphene flakes in different liquid environments and also in the film[49, 55]. The measured value of two photon absorption coefficient[55, 67] is, $\beta \sim 5 \times 10^{-8}$cm/W. Because of the saturable absorption property, graphene is utilized as a saturable absorber in an ultrafast mode-locked fiber laser to generate ultrashort laser pulses with duration of ~750 fs at wavelength of 1565 nm[21]. Not only the saturation absorption property but also the carrier-carrier scattering, the emission of optical phonons, the intra- and inter-valley carrier phonon scattering and the conversion into low-energy acoustic phonons via anharmonic decay, all these processes are also well studies by appropriate degenerate and non-degenerate pump-probe experiments [55, 65, 68, 69].

## 3.2 Probing the low energy excitations in carbon nanotubes by time resolved spectroscopy

In condensed matter, various low energy collective excitations occur in a very short time scale. Some of the examples are photons, plasmons, magnons, and superconducting energy gaps. The characteristic time scale for the associated dynamical phenomena occurs in ps and sub-ps. Optical phonons often strongly couple with electrons and participate in all energy and phase relaxation processes for electrons in solids. Non-perturbative approach to strong electron-optical phonon coupling in SWNTs is interesting, which is believed to be responsible for the current saturation behaviors in electronic transport as well as for the appearance of a broad and red-shifted Raman feature. Dynamical quantities of phonons, i.e., the lifetimes and dephasing times are the key parameters that characterize the related processes. These true life times can be captured in real time by femtosecond time-resolved pump-probe spectroscopy, which also helps in determining the electron phonon coupling strength.

Coherent phonon spectroscopic studies on SWCNTs have been reported in the literature[70-72] using spectrally-resolved and time-resolved ultrafast spectroscopy. A strong pump pulse initiates coherent lattice vibrations, and then a delayed, spectrally broad probe pulse induces additional lattice vibrations through impulsive stimulated Raman scattering. Figure 7(a) shows coherent generation of time-domain transmission modulations from a SWNTs sample measured at probe photon energy of 1.48 eV (840 nm). The oscillatory signal, which originates from the coherent lattice vibrations excited by the pump pulse, consists of high frequency and low-frequency contributions. In the Fourier transformed spectra in Fig. 7(b), the low-frequency peaks around 7 THz correspond to the radial breathing mode (RBM), while the G-mode of phonons appears at high frequency of 47.69 THz. The existence of multiple RBM peaks indicates that the sample contains SWNTs of several chiralities that are coherently and resonantly excited at 840 nm. As the spectral window for the probe pulse is changed, the G-mode phonons of the SWNTs show drastic changes with the probe wavelength in both amplitude and phase (see Fig. 7c). The coherent oscillatory signal gets almost completely suppressed when the probe energy is close to the center of the laser spectrum, i.e., 1.57 eV, while strong oscillations are observed at 1.65 eV and 1.46 eV, which are each separated from the center energy by roughly one-half of the G-mode phonon energy (~100 meV). Additionally, the G-mode amplitude decays monotonically with time delay for each probe photon energy, while the signal decay more rapidly at the lowest probe energy compared to the higher probe energy (Fig. 7c).

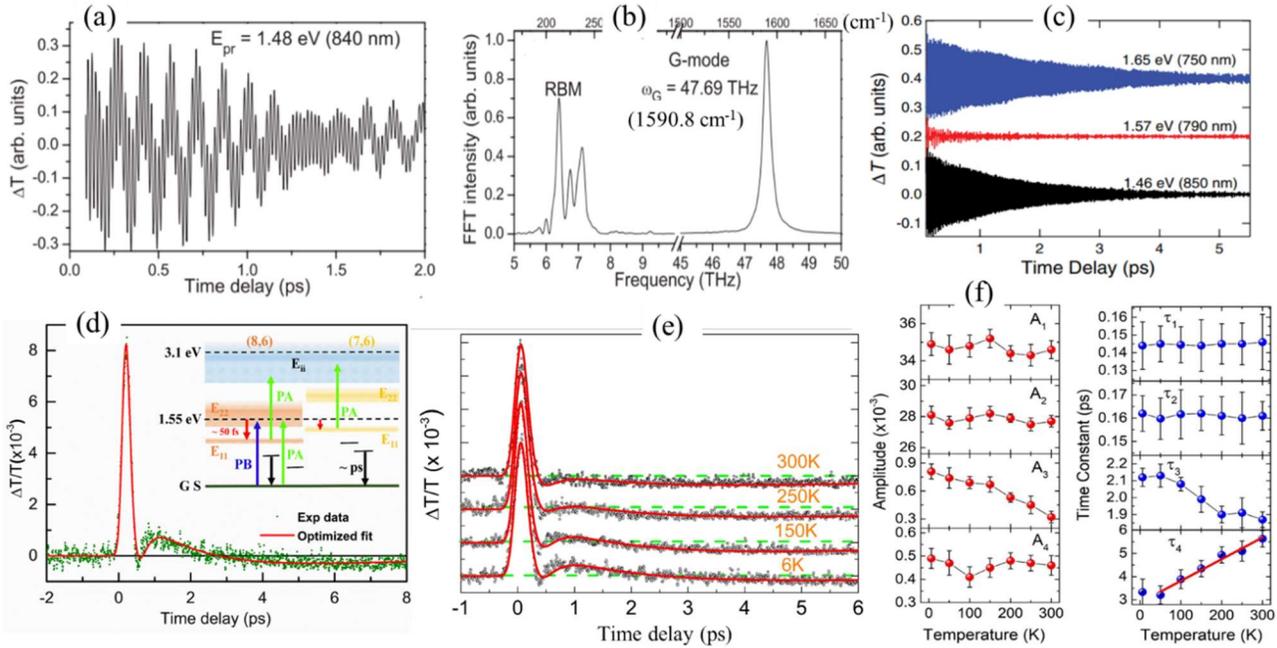

**Figure 7:** (a-c) Coherent excitation of phonons and (d-g) temperature dependence of the carrier relaxation dynamics in SWNTs. (a) Time-resolved pump-probe signal for pump and probe photon energies of 1.55 eV (800 nm) and 1.48 eV (840 nm), respectively, and (b) the corresponding Fourier-transformed spectrum showing coherent excitation of radial-breathing modes (RBMs) at 6.0-7.5 THz (200-250 cm$^{-1}$) of different chirality tubes, and the optical phonon mode (G-mode) at 47.69 THz (1590.8 cm$^{-1}$). (c) Effect of the probe photon energy as indicated on the coherent phonon dynamics by keeping the pump photon energy same at 1.55 eV (800 nm). Adapted from ref. [70]. (d) Experimentally measured time-resolved differential transmission from SWNTs film at 6 K. The continuous curve (in red) is a numerical fit obtained by using multi-exponential decay function convoluted with a Gaussian pulse (80 fs FWHM). Inset: Summary of the relaxation processes involved in the decay of transient response from the SWNTs. (e) Temperature dependence of transient transmission response from SWNTs obtained at constant pump fluence of 512 μJ/cm$^2$, and (f) evolution of the kinetic parameters: A's are amplitudes and τ's are time-constants of various relaxation components. Adapted from ref. [46].

We have recently studied temperature dependence of ultrafast photoresponse and signatures of low energy π–plasmons in SWNTs in a film (see Figs. 7(d-f)). The effect of the dark excitons and π-plasmons could be observed in the hot carrier relaxation dynamics, both of which showed a clear dependence on the sample temperature. Competing photo-bleaching and induced-absorption effects are observed simultaneously within the first 200 fs in the recovery of the photoinduced changes in the sample transmission involving processes of intraband relaxation and intertube coupling by charge transfer[46]. An early time evolution of the superimposed transient bleaching and induced photo absorption of almost similar strengths occurs, whereas at longer times, the photoresponse it is governed by slow recovery of long-lived dark excitons. After about 3 ps, the signal is dictated by the slowest negative relaxation component attributed to the low-energy π-plasmons (Fig. 7d). In the raw data, an absorption trough near 500 fs evolves with the increasing sample temperature. This particular feature is masked by the reduced induced transmission at room temperature and above. The π-plasmon dynamics is dictated by the interaction with the participating phonon population. Significant temperature dependence is observed in the related time constant, $\tau_4$ in Fig. 7(e). From the linear temperature dependence of the decay time constant of the π-plasmons, we have extracted the value of the electron−phonon coupling strength to be ∼0.86.

## 4. Terahertz time-domain spectroscopy and transient photoconductivity in graphene and carbon nanotubes

Time-domain THz spectroscopy is another technique where femtosecond laser pulses are used for generation and detection of coherent THz radiation. This technique using coherent THz radiation has numerous advantages, including the high sensitivity to the response of charge carriers, it covers the relevant energy ranges of exciton-binding energy, intraband transitions, and phonon resonances[73]. Most importantly, THz time-domain spectroscopy employing the THz pulses has the ability to extract the frequency-resolved complex conductivity and dielectric response, simultaneously. Now the experiments can be done with an optical excitation pulse and the induced changes at the THz frequencies can be captured in a time-resolved manner. A typical experimental layout is shown in Fig. 8, which can be used for just the time-domain THz spectroscopic measurements to extract the absorption and refraction properties at THz frequencies and also for time-resolved optical pump-THz probe spectroscopy. In the optical pump-THz probe experiments, what is effectively measured is the optical pump induced changes in the electric field associated with the THz probe pulse and monitor the change with time-delay between the optical pump and the THz probe pulses. The 2D spectroscopy involving time-resolved measurements for all the times in the THz pulse is quite time-taking and what is usually done is that one simply records the optical pump-induced changes in the peak of the THz probe pulse with respect the time-delay between pump and probe pulses. Here, the time of arrival of the pump-pulse onto the sample is being controlled. The pump-induced change in the peak of THz pulse electric field E(t), either in the reflection or the transmission, is the mean of the complete spectral weight carried within the THz probe pulse, and by doing so, one gets temporal evolution of the optical pump induced THz response of the sample under study. Generally, the change in peak amplitude as a function of pump pulse is proportional to the change in the conductivity of the sample, as probed with the full bandwidth of the spectrometer, and allows one to extract the free carrier kinetics in the system. For example, a decrease in broadband transmission would result from the formation of free carriers (e.g., charge injection), and the subsequent recovery back toward the baseline would result from carrier trapping or recombination. These measurements have sub picosecond time resolution and a time range of a few nanoseconds, limited by the length of the mechanical pump delay stage for pump pulse. Finally, the pump pulse can also be fixed at a specific delay time of interest, and probe pulse is scanned to measure the time-domain THz waveforms after that specific time of photoexcitation. The measured time or frequency-resolved THz response data from a metallic samples can be analyzed using a simple classical Drude-like model to gather information about the carrier mobility and charge density[74]. There are two main effects that contribute to the interaction with the incident wave at THz frequencies, i.e., the high-frequency conductivity associated with free charges (electrons and/or holes) and the dielectric effects caused by bound charges[75, 76].

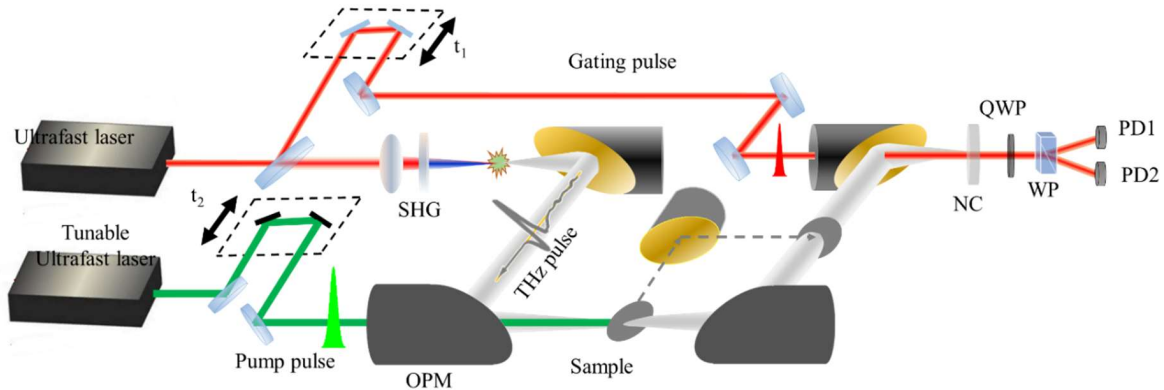

**Figure 8**: Typical experimental layout for THz time-domain spectroscopy and optical pump-THz probe time-resolved spectroscopy in both the reflection and transmission configurations. SHG: second harmonic generation crystal, OPM: off-axis parabolic mirror, NC: nonlinear crystal for electro-optic sampling, QWP: quarter wave plate, WP: Wollaston prism, PD: photodiode.

Measuring interesting physical phenomena requires an instrument with both sufficient spectral bandwidth and a high enough signal-to-noise ratio (SNR). Both, the bandwidth and SNR of a THz time-domain spectrometer are mainly decided by the THz generation and detection methods used. Optical rectification (OR) and electro-optical sampling (EOS) are the the most commonly amplifier-based solid-state THz generation and detection techniques, respectively. Difference-frequency analogue of second harmonic generation is the underlying process in OR, while EOS is based on the Pockels effect. ZnTe and GaP are the most popular choices for the above, with ZnTe having a relatively larger nonlinear coefficient and phase matching at 800 nm, and GaP having greater usable bandwidth (>7 THz)[75, 77]. LiNbO$_3$ (LNB)is also sometimes used to generate high-field THz pulses which can be useful in measuring THz nonlinearities or used as a strong THz pump source. LiNbO$_3$ (LNB) is also used as a strong THz pump source sometimes for generating high-field THz pulses which can be later useful in THz nonlinearities measurements. However, tilted-pulse-front pumping geometry required to satisfy phase matching conditions and the relatively narrow bandwidth (~0.1 to 2 THz) makes LNB not as common choice[75]. Air-based systems utilizing a combination of air plasma generation and air-biased coherent detection ABCD), commonly yield bandwidths in the range of 0.3 THz to >30 THz[78]. Therefore, they allow ultra-broadband "multi-THz" spectroscopy to be performed. Here, air plasma generation is fairly straightforward and cost-effective to implement, ABCD is much more challenging to construct and requires high voltages (tens of kilovolts)[78]. Therefore, many THz experiments utilizing air-plasma based THz source still employ EOS for detection. Photoconductive antennas are also commonly used for both generation and detection. However, they are used mostly used in high repetition rate ultrafast oscillator-based THz time-domain spectrometers. For generation, an external bias is applied to the emitter photoconductive switch, and the ultrafast current upon laser excitation emits THz radiation. The inverted effect can be used to detect THz electrical fields on another similar photoconductive switch. More recently, researchers have developed new schemes for generation of high power and broadband THz radiation. For example, the spintronics based THz emitters in which invers spin Hall effect is underplay in the heavy metal layer of the spintronic metallic thin film heterostructure[79, 80].

**4.1 Negative transient THz photoconductivity in graphene**

Ultrafast carrier dynamics in graphene have been extensively studied using optical pump–optical probe spectroscopy[49, 54, 81]. However, the high photon energy of optical probe limits the investigations largely to interband dynamics. To investigate the intraband response, THz photons are preferable. For the first time, George *et al*.,[82] reported photoexcited charge carrier dynamics in few atomic layer thick graphitic thin film by optical pump–THz probe spectroscopy. In their experiments, the sample was photoexcited by an ultrafast laser at 760 nm wavelength and the probe THz pulses were in 0.3-3.0 THz bandwidth. Figure 9(a) shows the measured changes of the peak THz transmission at the excitation pulse energy of 14.8 nJ. The decay kinetics relaxes via two pathways, i.e., rapidly decreasing process with a time-constant of ~1 ps and a slowly increasing process between 1 to 15 ps. The possible relaxation processes are schematically illustrated in Fig. 9(b). Soon after the photoexcitation, the non-resonantly excited electrons and holes are nonthermally distributed from the band edge. A hot Fermi–Dirac distribution is subsequently formed within 10-150 fs. Following this rapid thermalization, the hot charge carriers are cooled down to the band edge within 150 fs and 1 ps through intraband phonon scattering. After cooling down the hot carriers to the band edge, the nonradiative electron–hole recombination process dominates the photoconductivity and the THz transmission within time between 1 and 15 ps. Jnawali *et al*.,[83] observed negative THz photo-conductivity in CVD-grown graphene after the optical excitation as shown in Fig. 9(c). The observed negative THz photoconductivity is explained by considering the low-frequency limit of Drude conductivity expressed as $\sigma = D/\pi\Gamma$. When the main effect of photoexcitation is to increase the scattering rate $\Gamma$, the photoinduced reduction in conductivity is expected. Before and after the photoexcitation the fractional changes in the parameters $D$ and $\Gamma$ were 0.3% and 3.1%, respectively. Thus, the prime reason for the negative change of photoconductivity is the increase of scattering rate. To further gain a deeper insight into the mechanism of negative photoconductivity, Tielrooij et. al.,[84] utilized optical pump-THz probe with variable pump wavelength to investigate the pathways contributing to the ultrafast energy relaxation of photocarriers in CVD-grown graphene monolayers (Figs. 9(d,e)). It was found that the negative photoconductivity is directly related to the generation of secondary hot carriers during the energy relaxation cascade process. Immediately after photoexcitation, the carrier–carrier scattering between photocarriers and carriers in the Fermi sea occurs.. This promotes carriers from below to above the Fermi level and creates secondary hot electrons. These secondary hot electrons thus provide a negative contribution to the transient photoconductivity since the momentum scattering time increases with carrier energy[84, 85]. The photocarrier relaxation in graphene is mostly governed by carrier–carrier scattering and phonon emission. The negative photoconductivity is more significant when graphene is pumped with larger photon energy (Figs. 9(d,e)). This clearly suggests that carrier–carrier scattering is the dominant relaxation pathway because energy relaxation via phonon emission is independent of photon energy. Photons with greater energy excite electrons to higher energy states and facilitate more events of electron– electron scattering during the relaxation cascade resulting in a hotter carrier distribution (Fig. 9(e)). This process increases the number of hot carriers but does not increase the number of total carriers in the conduction band. Therefore, the change in the conductivity has a negative sign.

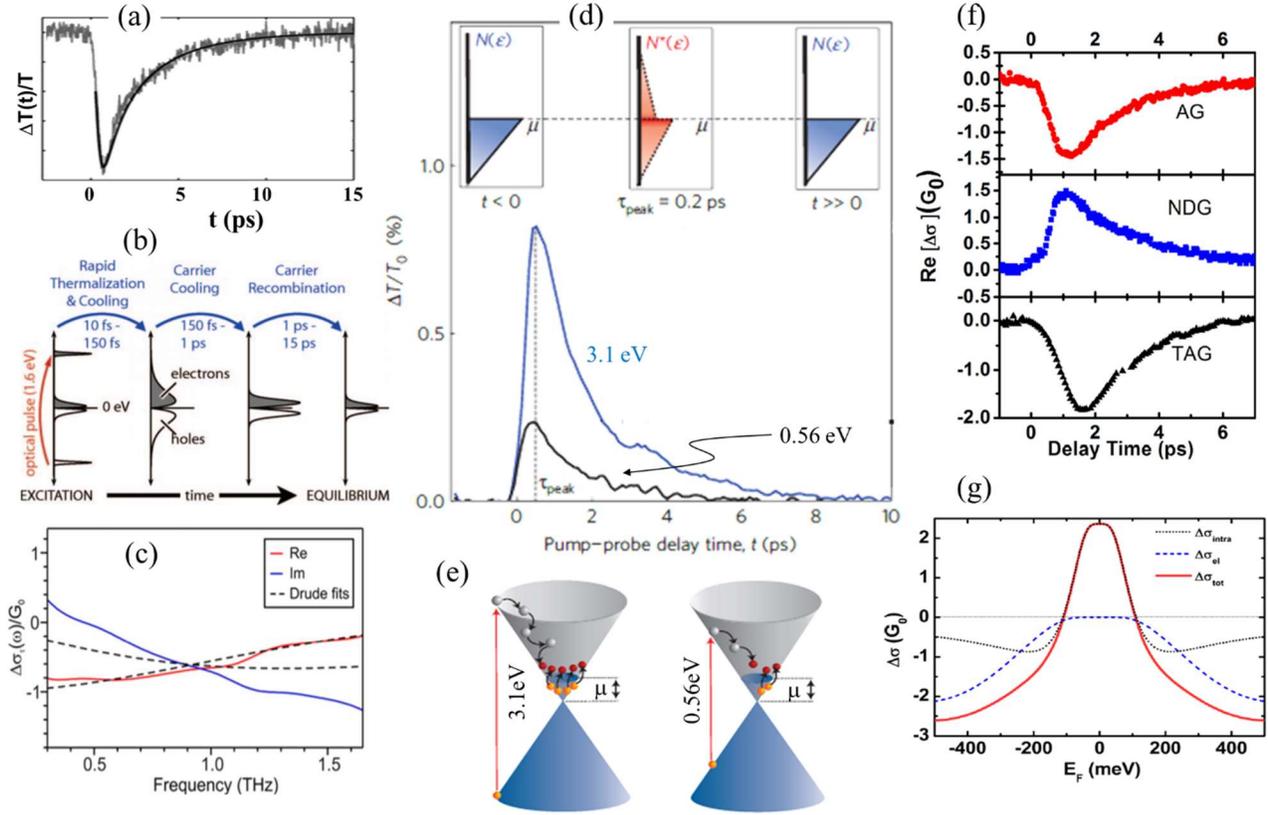

**Figure 9**: Optical pump-induced THz transient photoconductivity in graphene and graphitic thin films. (a) THz differential transmission recovery dynamics in ultra-thin graphitic film containing just 12 layers obtained by optical pumping at 1.6 eV, and (b) a schematic of the likely processes by which optically excited, non-equilibrium electron and hole distributions approach equilibrium. Adapted from ref. [82]. (c) Real and imaginary parts of the optical pump (1.5 eV) induced THz photoconductivity of graphene extracted from a data corresponding to a time of ~2.5 ps after the optical pulse excitation. Reproduced with permission from ref. [83]. (d) Induced THz differential transmission in graphene at two pump photon energies as indicated, and (e) interpretation of the possible relaxation processes in the Dirac cones. For a given absorbed photon density, a higher photon energy generates a higher signal, indicating a hotter carrier distribution. Adapted from ref. [84]. (f) THz differential transmission dynamics obtained by optical pumping at 1.58 eV in three types of graphene-like samples, the as-grown graphene monolayer, nitrogen-doped graphene monolayer and thermally annealed nitrogen-doped graphene monolayer, and (g) the numerically simulated change in THz conductivity as a function of Fermi energy in these three types of graphene-like samples. Adapted from ref. [86].

The negative photoconductivity of graphene can be facilitated via the pumping of photons over a wide range of wavelengths (from UV to IR). This virtue makes graphene to be more advantageous in optoelectronics than traditional semiconductors, where the photo response range is restricted by the bandgap energy. In addition to the secondary hot carrier generation, intraband scattering has been found to be an important contributor to negative photoconductivity in graphene. The magnitude and sign of the transient photoconductivity in graphene have been demonstrated to be determined by the relative contributions of the secondary hot carrier generation and intraband scattering, which depends on the Fermi energy. Kar *et al.*,[86] employed three samples including as-grown graphene monolayer (AG), nitrogen-doped graphene monolayer (NDG) and thermally annealed nitrogen-doped graphene monolayer (TAG) to represent the graphene monolayers with different Fermi energies. Transient photoconductivity is negative for the as-grown sample, while it is positive for the nitrogen-doped sample. After thermal annealing, the photoconductivity becomes negative again (Fig. 9(f)). The scattering of photocarriers in graphene has two dominant channels, the intraband scattering and the Coulomb scattering involving secondary hot carrier generation. The contributions of these two channels to the photoconductivity are related to Fermi energy. Therefore, the sign and magnitude of the photoconductivity is different for the three samples depending on the value of Fermi energy ($E_F$) (Fig. 9(g)). The $E_F$ of ~−180 meV and τ of ~34 fs facilitates negative photoconductivity in the as-grown sample, while the shift of $E_F$ towards the Dirac point results in positive photoconductivity in the nitrogen doped sample. The increased $E_F$ in the annealed sample makes the photoconductivity become negative again.

## 4.2 Stationary and photo-induced transient THz conductivity in carbon nanotubes

The characteristic frequencies of phonons, polarons, and excitons in solids, all lie in the THz frequency range giving THz spectroscopy access to study these important phenomena. Electron-phonon interactions often reveal intrinsic factors that limit the mobility and charge transport in a material system. At room temperature, phonon modes in the THz regime are thermally populated (298 K ~ 25.7 meV ~ 6.2 THz), meaning that electron−phonon interactions in the THz regime can have a significant effect on even the static conductivity. The dielectric response of CNTs has been investigated in detail using the THz time-domain spectroscopy to reveal a few interesting properties of the low energy electrons, the phonons and excitons. Frequency-dependent real and imaginary parts of the complex dielectric function as shown in Fig. 10(a), were measured for SWNTs dispersed in a poly(vinyl) alcohol host matrix using the time-domain THz spectroscopy. The low-frequency phonons of carbon nanotubes which were theoretically predicted but not observed directly in experiments, could be observed here[87], for the first time, at frequencies of ~0.26, 0.60, and 0.85 THz (Fig. 10(a)). The low-frequency flexural modes of CNTs could also be observed for DWNTs in a freely standing thin film[88].The frequency-dependent real index of refraction and extinction coefficient, and hence the complex dielectric function of the film of thickness ~200 nm is presented in Fig. 10(b) that were obtained by using time-domain THz spectroscopy in the frequency range 0.1 to 2.5 THz. The real index of refraction and extinction coefficient have very high values of approximately 52 and 35, respectively, at 0.1 THz, which decrease at higher frequencies. The low frequency phonon modes of the carbon nanotubes at ~0.45 and 0.75 THz are clearly seen in the real and imaginary parts of the complex dielectric function riding on a broad background resonance centered at around 1.45 THz. The latter being similar to that in SWNTs was assigned to low-energy electronic excitations arising from curvature and bundling effects induced band gap opening. These experiments also suggested a possible application of DWNTs free standing films as a neutral density filter in the THz range.

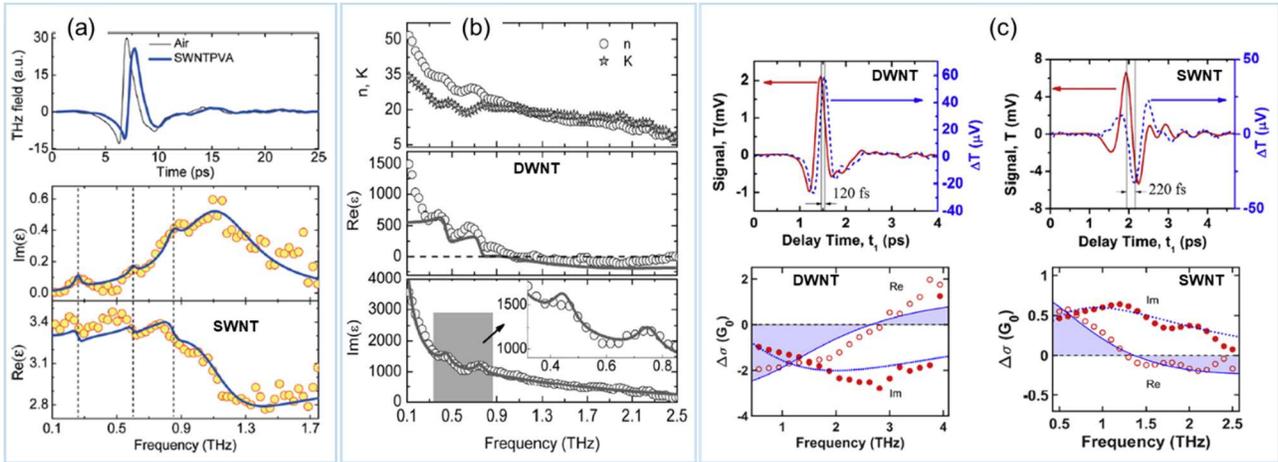

**Figure 10:** Static and dynamic THz characteristics of carbon nanotubes in a host medium or suspended without a substrate. (a) Top: The electric field of the THz pulse as recorded with and without a sample containing SWNTs dispersed in a PVA film, bottom: imaginary, and real parts of the complex dielectric function. The vertical dashed lines locate the resonance frequencies arising due to low frequency phonon modes of SWNTs. Adapted from ref. [87]. (b) Real index of refraction and extinction coefficient (top), and the Re(ε) and Im(ε) parts of the complex dielectric function (bottom), extracted from THz time-domain signal recorded in air and with a DWNT film. Various phonon resonance can also be seen. Adapted from ref. [87]. Optical pump (1.5 eV) induced THz transmission changes (top) and the corresponding photoconductivity spectra after certain time of photoexcitation for DWNT and SWNT films. Adapted from ref. [89].

The low photocarrier injection in conventional semiconductors enhances the conductivity because of increased intraband absorption at THz radiation frequencies. Kar *et al.*,[89] reported frequency-dependent photoconductivity in THz range after 800 nm optical pump excitation in (6,5) semiconducting SWNTs sample and DWNTs sample containing both metallic and semiconducting tubes. The real and imaginary parts of photoconductivity show non-Drude behavior (see Fig. 10(c)). In case of SWNTs, the real part of THz photoconductivity is positive at low frequencies while it becomes negative at the high frequencies. In contrast, the trend with respect to the frequency gets reversed for the DWNTs film, real part of the photoconductivity is negative at low frequencies and it's positive at the high frequencies. This contrasting behavior is explained using Boltzmann transport theory, where the carrier scattering rate is energy-dependent. Taking the scattering rate to be dominated by short-range disorder scattering, it was shown[89] that the Boltzmann transport model captures most of the experimental observation in Fig. 10(c).

## 5. Conclusions and future perspectives

Tremendous advances have already been made in the area of the nanocarbon science and technology as a consequence of breakthroughs in synthetic strategies and advancement in various optical measurement methodologies for their investigations. However, the research trends clearly indicate that there is still lot more to be done in terms of achieving full understanding of these materials and their vast potential applications. Thanks to the advanced time-resolved techniques, in particular, the pump-probe spectroscopy and THz time domain spectroscopy, researchers have started using them in new and more innovative ways to explore and unearth the unhidden, sometimes, by creating extreme experimental conditions of temperature, pressure, etc. Further studies are definitely required in these directions. As indicated by a few studies covered in this article, graphene and carbon nanotubes have significant virtues for use in ultrafast and THz devices such as the sources, detectors and modulators. Ultrafast dynamical studies always make an essential part in the investigations on such materials for Deep insights into the interlayer transient behavior of charge carriers in complex device structures and hence to provide guidance for the design and fabrication of more efficient structures and devices. Besides of the excellent broadband saturable absorption and ultrashort recovery time in graphene, the long-term stability of these 2D sheets should also be further investigated and developed.

**Acknowledgement**

SK acknowledges Department of Science and Technology, Govt. of India for financial assistance. AS acknowledges Council of Scientific and Industrial Research for Senior Research Fellowship.